\documentclass[preprint,epsfig,onecolumn,floats,showpacs]{revtex4}
\usepackage{amssymb}
\usepackage{graphicx}
\usepackage{epsfig}
\usepackage{amsmath}
\usepackage{amsfonts}
\usepackage{dcolumn}
\usepackage{bm}
\usepackage[dvipdfmx]{hyperref}

\begin{document}

\title{Experimental Studies of Scaling Behavior of a Quantum Hall System
with a Tunable Landau Level Mixing}
\author{Y. J. Zhao$^{(1)}$}
\author{T. Tu$^{(1)}$}
\email{tutao@ustc.edu.cn}
\author{X. J. Hao$^{(1)}$}
\author{G. C. Guo$^{(1)}$}
\author{H. W. Jiang$^{(2)}$}
\author{G. P. Guo$^{(1)}$}
\email{gpguo@ustc.edu.cn}
\affiliation{$^{(1)}$ Key Laboratory of Quantum Information, University of Science and
Technology of China, Chinese Academy of Sciences, Hefei 230026, P. R. China\\
$^{(2)}$ Department of Physics and Astronomy, University of California at
Los Angeles, 405 Hilgard Avenue, Los Angeles, CA 90095, USA }
\date{\today}

\begin{abstract}
Temperature dependence of the longitudinal and Hall resistance is studied in
the regime of localization-delocalization transition. We carry out
measurements of a scaling exponent $\kappa$ in the Landau level mixing
region at several filling factors. The localization exponent $\gamma$ is
extracted using an approach based on the variable range hopping theory. The
values of $\gamma$ and $\kappa$ are found to be universal, independent of
filling factor in our sample. We can conclude that although Landau level
mixing can change the degeneracy of a quantum Hall state, the value of the
scaling exponent remains the same for a given sample that contains a fixed
disorder profile.
\end{abstract}

\maketitle

\pagestyle{headings}


The localization problem has been studied intensively in the past decades
\cite{Huckestein1995,sondhi1997}. It's well accepted that there is no true
delocalization state in non-interaction two-dimensional electron gas (2DEG)
provided with disorder. However when a strong external magnetic field is
applied, a delocalized state is formed at the center of the Landau level,
while the electronic state is localized away from this discrete energies.
The coexistence of localized and delocalized states is essential for the
quantum Hall effect. Phase transitions between localized and delocalized
states will occur when the Fermi level is swept from one Landau level to
another. The relative position of the Fermi level and Landau levels can be
experimentally changed by varying the external magnetic field and/or the
carrier density. Since this phase transition is considered as a continuous
quantum phase transition, finite size scaling theory \cite{Pruisken1988} can
be applied in the critical regime, where the resistance tensor scales as $%
R_{uv}=R_{c}f(L/\xi )$ for a sample of finite size L. Here $f$ is a scaling
function which can be derived from the microscopic calculation. The
localization length $\xi $ diverges as the Fermi level approaches the center
of a Landau level $E_{c}$ as $\xi \propto \left\vert E-E_{c}\right\vert
^{-\gamma }$ with an exponent $\gamma $. The interaction induced quantum
phase coherence length sets the effective sample size $L$ and its
temperature dependence of $L\propto T^{-\frac{p}{2}}$. Then one obtains $%
R_{uv}=R_{c}f(\left\vert B-B_{c}\right\vert T^{-\kappa })$, the scaling
function of both the longitudinal resistance $R_{xx}$ and the Hall
resistance $R_{xy}$, where the scaling exponent is expressed as $\kappa
=p/2\gamma $. Approaching zero temperature, the maximum slope in the Hall
resistance $R_{xy}$ with varying magnetic field $B$ diverges as a power law $%
\frac{dR_{xy}}{dB}\mid _{B_{c}}\propto T^{-\kappa }$, while the half width
for the longitudinal resistance $R_{xx}$ vanishes as $\Delta B\propto
T^{\kappa }$ \cite{Wei1988}.

Despite the intense studies of the scaling behaviors of the quantum Hall
systems \cite%
{Wei1988,Koch1991b,Koch1991,Wei1992,Engel1993,Haug2001,Haug2002,Haug2002b,Li2005}
, several issues are still unsettled. For example, when the electron spin is
unresolved (i.e., the two spin states of the Landau levels are mixed), the
interaction of electrons in different Landau levels complicate the problem
and draw into question whether this universality of scaling behavior can be
preserved. Experimentally, there was an indication that the exponent $\kappa
$ can be changed by a factor of two when the system becomes spin-degenerate
\cite{Wei1990,Wei1993}. On the other hand, theoretical works \cite%
{Wen1994,Lee1994}, based on two different models, concluded that Landau
level mixing will not change the universality of this phase transition. The
earlier experimental studies on the localization-delocalization phase
transition in the Landau level mixing regime is focused on the spin
unresolved plateau transitions \cite{Wei1990,Wei1993}, which involves the
adjacent Landau levels with spin degeneracy. The experiment presented in
this paper is carried out in a two subband system in which two Landau levels
(LLs) with different subband, Landau level index, or spin state can be
controllably mixed either by varying electron density or by changing
magnetic field. The main objective of the study is to determine
experimentally the contribution of the Landau level mixing on the scaling
behavior.

The sample we investigated is grown by molecular-beam epitaxy and consists
of a symmetrical modulation-doped $24$ nm wide single GaAs quantum well
bounded on each side by Si $\delta $-doped layers of AlGaAs. Heavy doping
creates a very dense 2DEG, resulting in the filling of two subbands in the
well when at low temperature. As determined from the Hall resistance data
and Shubnikov-de Haas oscillations in the longitudinal resistance, the total
density is $n=8.0\times 10^{11}$ cm$^{-2}$, where the first and the second
subband have a density of $n_{1}=6.1\times 10^{11}$ cm$^{-2}$ and $%
n_{2}=1.9\times 10^{11} $ cm$^{-2}$. The sample has a low-temperature
mobility $\mu =4.1\times 10^{5} $ cm$^{2}$/V s, which is extremely high for
a 2DEG with two filled subbands. The samples are patterned into Hall bars
using standard lithography techniques. A NiCr top gate is evaporated on the
top of the sample, approximately $350$ nm away from the center of the
quantum well. By applying a negative gate voltage on the NiCr top gate, the
electron density can be tuned continuously. Magneto-transport measurements
were carried out in an Oxford Top-Loading Dilution Refrigerator with a base
temperature of $15 $ mK. To measure the longitudinal and Hall resistance $%
R_{xx}$ and $R_{xy} $, we used a standard ac lock-in technique with electric
current ranging from $10$ nA to $100$ nA at a frequency of $11.3$ Hz.

\begin{figure}[htbp]
\includegraphics[width=1.0\columnwidth]{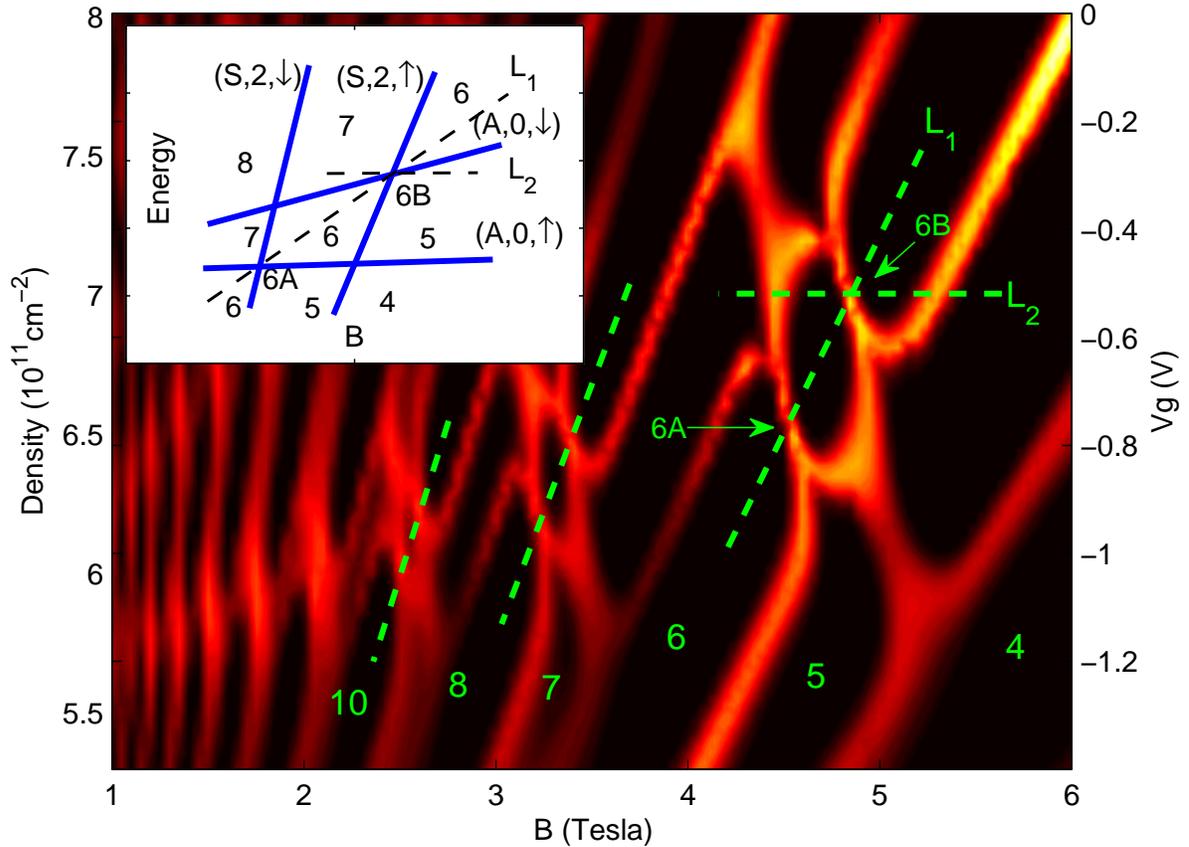}
\caption{(color online). The phase diagram of $R_{xx}$ against gate voltage $%
V_{g}$ and magnetic field $B$. Numbers in the graph label the regions with
corresponding filling factor. The horizontal dashed line $L_{2}$ is the
conventional measurement trace of plateau transition between filling factor
5-6-7, while the three tilted dashed lines are measurement path at filling
factor 6, 8, 10 separately. Especially at the region of filling factor $%
\protect\nu=6$, 6A and 6B indicate the Landau level mixing points where the $%
R_{xx}$ peaks occur. Inset is the Landau level diagram at $\protect\nu=6$.}
\label{phase}
\end{figure}

At first we measure the phase diagram of the two-subband sample, which is
the gray-scale plot of longitudinal resistance $R_{xx}$ as a function of
magnetic field $B$ and electron density $n$, as shown in Fig.\ref{phase}. In
this gray-scale map, bright lines represent peaks in $R_{xx}$, where the
electron state is delocalized; while the dark regions are the minimums,
corresponding to the quantum Hall states, where localized state occurs. From
this diagram we can find the spin is resolved when magnetic field is greater
than 2.2 Tesla. The ringlike structures at even filling factor is due to
interaction between the two set of Landau levels \cite{Jiang2005}. This
situation occurs when two Landau levels from different subband are brought
into degeneracy or mixing by tuning electron density and magnetic field.
According to the standard Landau level fan diagram as illustrated
schematically in Fig.\ref{phase} inset, point 6A represents the mixing of
Landau level $(A,0,\uparrow )$ and $(S,2,\downarrow )$, while point 6B
corresponds to mixing of $(A,0,\downarrow )$ and $(S,2,\uparrow )$. Here we
label the single-particle levels $(i,N,\sigma )$, and $i(=S,A)$, $N$, and $%
\sigma (=\uparrow ,\downarrow )$ are the subband (symmetry or antisymmetry),
orbital and spin quantum numbers. Since Landau level mixing influence the
pattern of phase diagram profoundly, it's interesting to find out whether
this mixing will change the universality of scaling behavior of
localization-delocalization transition or not.

\begin{figure}[tbph]
\includegraphics[width=1.0\columnwidth]{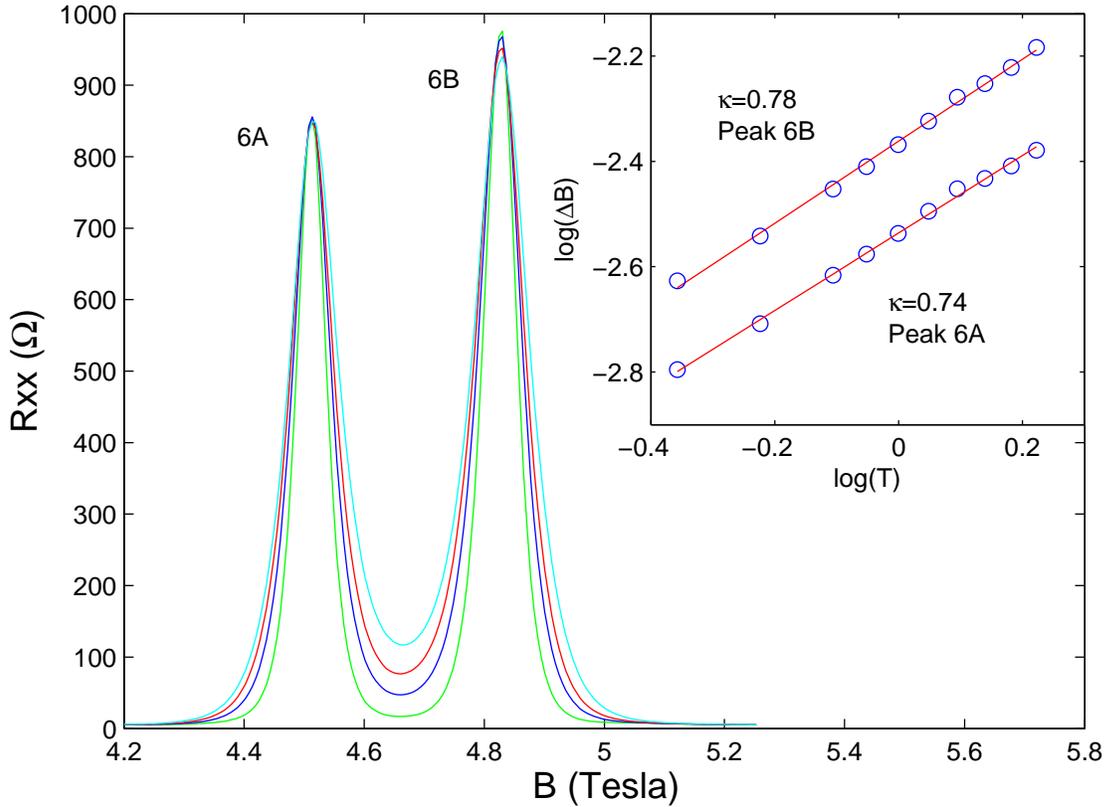}
\caption{(color online). The longitudinal resistance $R_{xx}$ at
different temperatures along trace $L_{1}$. The insets shows the
half-width $\Delta B$ vs temperature $T$ corresponding to the two
peaks in the $R_{xx}$ curve.} \label{Sigmaxx-T}
\end{figure}

\begin{figure}[tbph]
\includegraphics[width=1.0\columnwidth]{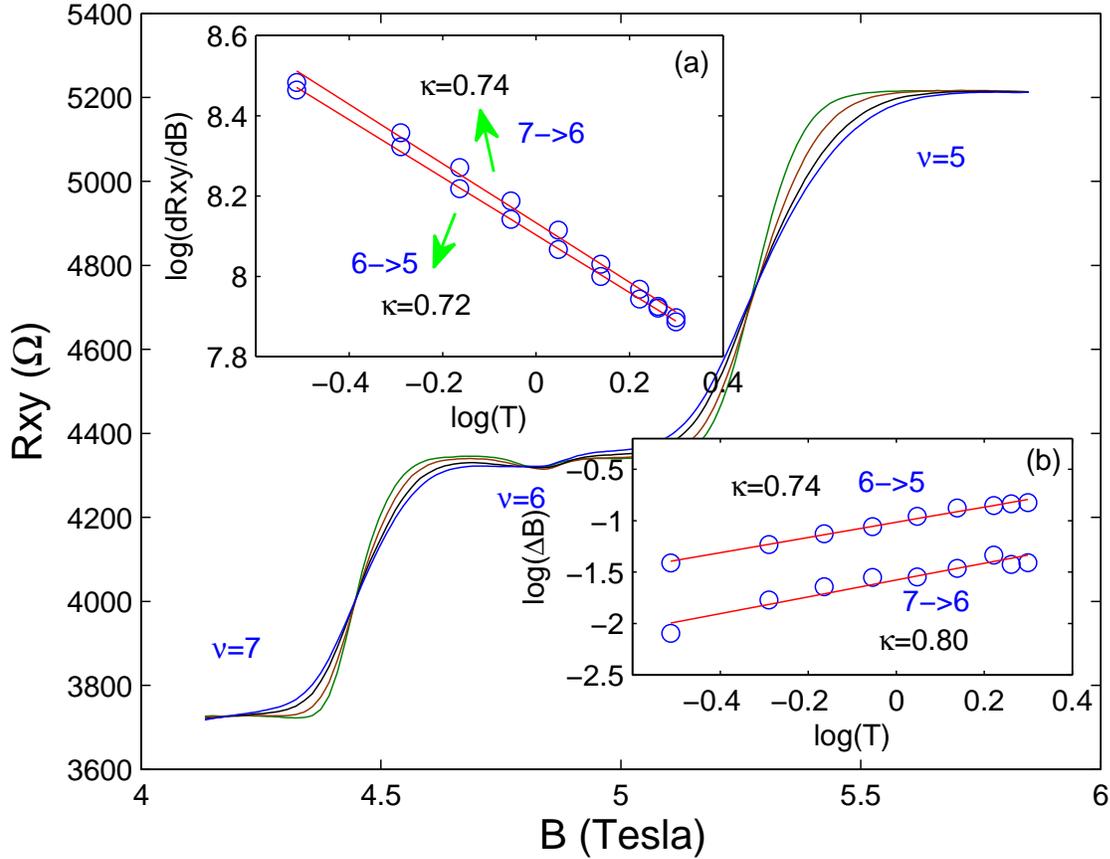}
\caption{(color online). The Hall resistance $R_{xy}$ at different
temperatures along line $L_{2}$. Inset (a) shows the scaling exponent $%
\protect\kappa $ extracted from the temperature dependence of $R_{xy}$ slop,
while inset (b) gives $\protect\kappa $ derived from scaling of half width
of $R_{xx}$ peak.}
\label{v-6-line-c}
\end{figure}

By changing the magnetic field whistle varying the gate voltage, we can
guide the measurement trace crossing bright lines in the phase diagram and
then undergoing localized-delocalized transitions. Dashed lines in Fig.\ref%
{phase} illustrate such measurement traces. It should emphasis that the
tilted traces crossing ringlike structures at filling factor $\nu =6,8,10$,
are important for our study since they cross the Landau level mixing points
exactly. Fig. \ref{Sigmaxx-T} shows the temperature dependence measurement
of longitudinal resistance $R_{xx}$ following trace $L_{1}$ across the
Landau level mixing region at filling factor $\nu =6$. The two peaks which
result from intersection of trace $L_{1}$ and the ringlike structure, are
noted as 6A and 6B, and each shows a broaden as the temperature increasing
from $50$mK to $1.2$K. In order to compare the experiment result with
conventional plateau-plateau transition situation, we also measured the
transition between filing factor $\nu =5,6,7$ along trace $L_{2}$. The
corresponding result is plotted in Fig.\ref{v-6-line-c}.

Temperature scaling exponent $\kappa $ can be extracted from power-law fit
for the half width $\Delta B\propto T^{\kappa }$, which yield values of $%
0.74 $ and $0.78$ for peak 6A and 6B respectively (Fig.
\ref{Sigmaxx-T} inset). It should be noticed that, along line
$L_{1}$ the Hall resistance keeps constant since the the measurement
trace will not change the filling factor. On the other hand, as
shown in Fig.\ref{v-6-line-c} inset, the scaling exponent is
extracted from half width $\Delta B$ as well as maxima of
$\frac{dR_{xy}}{dB}$, which shows $\kappa $ ranges within $\kappa
=0.75\pm 0.05$. The result is consistent with what occurs at Landau
level mixing region within experiment error. The scaling in the
Landau level mixing region at other filling factor ($\nu =8,10$) are
also examined. All the results fall into range $\kappa =0.75\pm
0.05$. Our experiment results turn out that the exponent $\kappa$ is
universal in respect to the filling factor within the experimental
error, whatever there exists Landau level mixing or not.

Additionally, an alternative approach to scaling can be used for a direct
evaluation of the localization length $\xi$. Polyakov and Shklovskii argued
that the mechanism for the conductivity peak broadening is variable range
hopping (VRH) in the presence of Coulomb interactions\cite{Polyakov1993}.
The scaling function of VRH conductivity is given as

\begin{equation}  \label{scaling-function}
\sigma_{xx}=\sigma_{0}\exp(-\sqrt{\frac{T_{0}}{T}}), T_{0}(B)\propto \frac{%
e^2}{\varepsilon\xi(B)}.
\end{equation}
The characteristic temperature $T_{0}$ is determined by the coulomb
interaction at length scale given by localization length $\xi$. Here $%
\sigma_{0}\propto 1/T$ is a temperature dependent parameter. As mentioned
above, the localization length $\xi(B)$ diverges as $\xi(B)\propto
|B-B_{c}|^{\gamma}$, combine this relations yields

\begin{equation}  \label{VRH-function}
\sigma_{xx}=\sigma^{\ast}s\exp(-\sqrt{T^{\ast}s}), s=\frac{|B-B_{c}|^{\gamma}%
}{T}.
\end{equation}
Where $s$ is the scaling variable, $\sigma^{\ast}$ and $T^{\ast}$ are
constant. Using this formula between conductivity $\sigma$ and $\xi$, we can
easily and directly obtain the localization length $\xi$ and corresponding
localization exponent $\gamma$.

\begin{figure}[tbph]
\includegraphics[width=1.0\columnwidth]{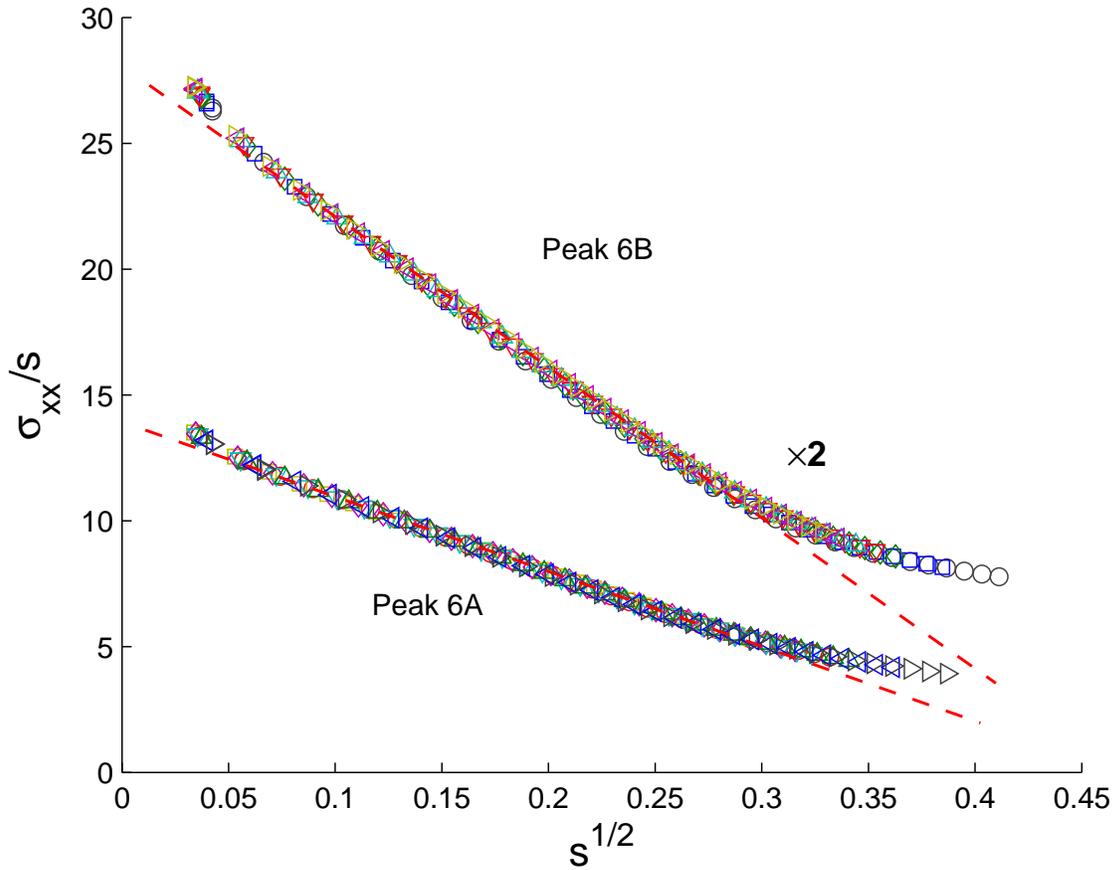}
\caption{(color online). Longitudinal conductivity $\protect\sigma _{xx}$ as
a function of the scaling variable $s$ in the transition regime at $\protect%
\nu =6$ along line $L_{1}$. The dashed line is indicated for fitting to Eq.(%
\protect\ref{VRH-function}).}
\label{fit-gamma}
\end{figure}

We fit the experiment data using the scaling function Eq.\ref{VRH-function}
as shown in Fig.\ref{fit-gamma}. By varying the localization length exponent
$\gamma $ we can make all the data in the phase transition region along line
$L_{1}$ fall into a single line. Therefore we find the fit is perfect, and
we get $\gamma \approx 1.3$. This result is quite different from that of
previous experimental \cite{Koch1991,Haug2001,Haug2002} and theoretical
works \cite{Chalker1988,Huckestein1990,Lee1993} which give a value $\gamma
\approx 2.3$. We also investigate the localization length exponent at other
region of the phase diagram, including Landau level mixing points at higher
filling factor ($\nu =8,10$), and conventional plateau-plateau transition
situation along line $L_{2}$ (between plateau $\nu =5,6,7$). Within
experimental error, the localization length exponent $\gamma $ is found to
be universal, independent of filling factor, Landau level mixing situation.
All the cases give a perfect fit of $\gamma =1.3\pm 0.2$.

The scaling exponent $\kappa $ and localization length exponent $\gamma $
both show a deviation from conventional values. Even these deviations,
however, we should emphasis that the universality of scaling behavior exist
for our sample, with exponent values of $\kappa =0.75\pm 0.05$ and $\gamma
=1.3\pm 0.2$. We can conclude that although Landau level mixing can change
the degeneracy of a quantum Hall state, this mixing will not change the
scaling behavior of our sample. Since the two exponents are extracted
independently, we think they reveal more valuable information beneath the
scaling phenomenon. Using the relation $\kappa =p/2\gamma $, we can get
inelastic scattering rate exponent $p\approx 2$, which is exactly the
zero-field clean limit for electron-electron scattering in two-dimensional
Fermi liquids \cite{Lee1993,Lee1996,Huckestein1999}. This result is also
consistent with most of the experimental works \cite{Engel1993,Haug2002b}.

For the samples we investigated, the deviation of $\kappa $ from the
conventional value $\kappa \approx 0.42$ can be attributed to the effect
from quantum localization toward classical percolation (using the value of $%
p $ and the $\kappa =p/2\gamma $ relationship, one can obtain
$\kappa \approx 0.42$ and $\kappa \approx 0.75$ corresponding to
$\gamma \approx 2.4$ and $\gamma \approx 1.3$ respectively). Most of
direct measurements of localization length exponent $\gamma $ yield
a universal value of $2.4$, which is predicted by theoretical
calculations based on a network model of quantum percolation
\cite{Chalker1988,Huckestein1990,Lee1993}. However, our
direct measurement of $\gamma $ through VRH method yields the value of $1.3$%
, which coincides with $\gamma =4/3$, a value obtained with theories of
classical percolation \cite{Lee1993,Trugman1983}. It shows that classical
percolation dominates scaling behavior in our samples. Theoretical work\cite%
{Moore2002} suggest that in presence of long range potential fluctuation due
to the remote ionized impurities in AlGaAs, a larger crossover region will
occur where classical percolation applies. On the other hand, in \cite%
{Li2005} the author provided a systematic investigation on the
influence of type of disorder potential on the scaling behavior. The
result turned out that only for short range alloy potential
fluctuations the scaling exponent is consistent with the
conventional value $\kappa \approx 0.42$. However for the long range
potential fluctuation situation, the exponent increase towards the
classical value of $0.75$. We think our results are coincide with
previous observations. This suggests that whatever there exists
Landau level mixing or not, the value of the scaling exponent
remains the same for a given sample that contains a fixed disorder
profile.

There is also a different approach for the explanation of this phenomena.
Avishai and Meir \cite{Meir2002} studied critical exponent $\gamma $ in the
presence of spin-orbit scattering. Their numerical calculations show that
the exponent is very close to $\gamma =4/3$ and consistent with our results.
The spin-orbit interaction is also investigated in GaAs/AlGaAs quantum well
recently \cite{Hirayama2005}, which suggested the this interaction arises
from the structural inversion asymmetry of the heterostructure, and hence is
influenced significantly by applying gate voltage. Since our sample
utilizing a front gate to tune the carrier density, it is possible for the
enhancement of spin-orbit interaction, which eventually induces localization
length exponent $\gamma \approx 1.3$.

In conclusion, temperature dependence of the Hall and longitudinal
resistance of two-subband quantum Hall samples is measured in our
experiments. The emphasis is focused on the scaling behavior of the
localized-delocalized transition in the Landau level mixing region at
several filling factors (from $5$ to $11$). The scaling exponent $\kappa $
is extracted by power law fit of half width of $R_{xx}$ peak dependence on
temperature. Additionally we extract quantitative information of the
localization exponent $\gamma $ through an approach based on the variable
range hopping theory. The main result is that we find an accurate
universality for both the scaling exponent $\kappa \approx 0.75$ and the
localization length exponent $\gamma \approx 1.3$ within the experimental
error. In comparison with results obtained from conventional plateau-plateau
transition, it turns out although Landau level mixing can change the
degeneracy of a quantum Hall state, the value of the scaling exponent
remains the same for a given sample that contains a fixed disorder profile.
Also several explanations are supposed to explain the value of $\gamma $ and
$\kappa $'s deviation from the conventional values.

This work was funded by National Fundamental Research Program, the
Innovation funds from Chinese Academy of Sciences, NCET-04-0587, and
National Natural Science Foundation of China (Grant No. 60121503, 10574126,
10604052).

\end{document}